# High precision time measurements in future experiments

J. Va'vra

SLAC National Accelerator Laboratory, CA 94025, USA

*Abstract* – Detector examples discussed: MRPCs, MCP-PMTs, Diamond detectors, SiPMTs, Low and high gain Avalanche diodes (LGADs) and Micromegas. We specifically discuss issues such as single pixel vs. multi-pixel tests, small test vs. large physics system results and hidden problems people usually do not want to talk about.[1]

## Content

This work supported by the Department of Energy, contract DEAC02-76SF00515.

[1] Invited talk at Micro-Pattern Gaseous Detector conference at La Rochelle, France, May 7, 2019, and INSTR20, Novosibirsk, Feb.26, 2020

# 1. Introduction

This paper is a follow up to a recent review paper [1]. This paper will update this review with new developments.

There is a general push for higher luminosity not only at LHC but also at Belle-II, Panda, Electron-ion collider, etc. Timing is more and more important. For example, ATLAS needs to connect charged tracks to the correct production vertices, using position resolution and timing resolution of ~30 ps/MIP [2]. Similarly, new DIRC applications aim for single photon timing resolution at a level of ~70-120 ps/photon. The new luminosity upgrades by a factor of 100-1000 compared to previous generation experiments will create a new demand on detectors. This means significant improvements are necessary in detector aging, electronics advances, and rate capability of detectors.

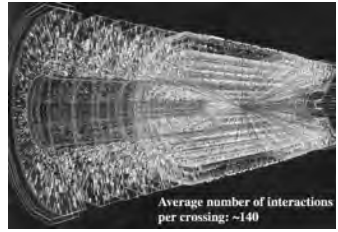

**Figure 1** ATLAS event after an LHC upgrade to a luminosity of ~$10^{35}$ cm$^{-2}$ sec$^{-1}$ (ATLAS collaboration).

# 2. Fast timing at a level of 20-30 ps per minimum ionizing particle (MIP)

A TOF technique seems simple, but there are many hidden effects which have to be overcome. There has been considerable progress with small single-pixel devices, however progress in very large systems has been slow. In the following I will discuss possible reasons.

One usually starts with a simple, somewhat naive formula in order to judge the timing resolution limit ($\sigma_{time}$):[2]

$$\sigma_{time} = \sigma_{noise}/(dS/dt)_{threshold} \sim t_{risetime}/(S/N), \qquad (1)$$

where S is the signal amplitude, N = $\sigma_{noise}$ is a noise. One can say: "Show me your pulses and noise level, and I will estimate your resolution." For a typical MCP-PMT risetime of $t_{risetime}$ ~ 200 ps, one needs S/N ~10 to get into the ~20 ps regime. For a slower Si-detector risetime of $t_{risetime}$ ~2 ns, one needs S/N ~100 to get into the same timing regime. However, there are many other contributions. First, we mention a few standard contributions due to electronics, chromatic effects, number of photoelectrons ($N_{pe}$), transit time spread (TTS), tracking, and $t_0$ time, which is experiment's start time (we neglect tracking effects as they depend on exact geometry):

$$(\sigma_{Total})^2 \sim [(\sigma_{Electronics})^2 + (\sigma_{Chromatic}/\sqrt{N_{pe}})^2 + (\sigma_{TTS}/\sqrt{N_{pe}})^2 + (\sigma_{Track})^2 + (\sigma_{to})^2 + \ldots] \qquad (2)$$

The $t_0$ time can easily dominate. There are many other obscure effects, which can affect timing resolution. For example: time walk, cross-talk effects in multi-pixel detectors, baseline ringing in multi-pixel detectors affecting later arriving pulses, charge sharing in multi-pixel detectors, chromatic effects, clock distribution throughout the system, pulse tail recovery, calibration, etc. The electronics resolution can reach ~2 ps in the good examples [1], while the best overall is the DRS4 electronics reaching <1 ps for short delay between start & stop [3].

## 2.1 MRPC

Figure 2a shows ALICE present MRPC TOF detector, which is now running about 15 years. It is delivering ~60 ps resolution and a maximum rate of ~500 Hz/cm$^2$ presently [4]. The most important design feature of this detector is a differential design, both on inputs and outputs, to minimize pick-up and cross-talk effects [4]. The detector uses the NINO ASIC with amplifier + discriminator + time-over-threshold (TOT) pulse height correction to time walk [5]. The NINO chip was a very significant instrumentation development in retrospect. Figure 2b shows a response of the NINO input stage to various input charges. It shows ~ 1ns peaking time independent of input charge. As a result of a well behaving design, the TOT correlation between time and width is very linear, as figure 2c shows; in addition, it has a very low power consumption (40 mW/channel).

The present plan for ALICE is to develop a new detector for high luminosity upgrade using a lower resistivity 400 μm-thick glass, allowing to build 20-gap MRPC capable of handling rates up to ~50 kHz/cm$^2$ with aim for ~20 ps resolution per MIP [4].

---

[2] Well known formula in the communication engineering.

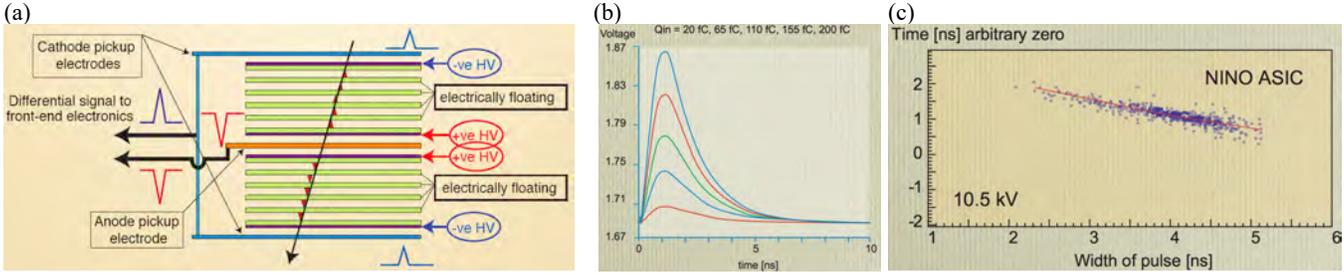

**Figure 2** (a) ALICE 10-gap MRPC TOF detector and its differential signal to front-end electronics [4]. (b) Response of the NINO ASIC input stage to various input charges from a pulser. One can see that pulse response is perfect allowing the time-over-threshold (TOT) pulse height correction [5]. (c) The correlation between time and width is very linear in this design [5].

## 3. MCP-PMT
### 3.1 Single-pixel MCP-PMTs

Microchannel plate detectors can achieve the best timing resolution presently. The Photek company measurements [6] demonstrated that the smaller pore size and higher MCP-to-anode electric field, faster the risetime. Figure 3 demonstrates this capability with an 18 GHz BW scope, indicating a 66 ps risetime for 3.2 µm pores and 95 ps for 6 µm pores using Photek-110 single pixel MCP-PMT [6]. This is the best demonstration of MCP-PMT timing capability to my knowledge. Using our simple equation 1, and neglecting all other contributions, one could get a single photon timing resolution of $\sigma_{TTS}$ ~3 ps if we assume signal-to-noise ratio S/N ~20, or better if S/N is higher.

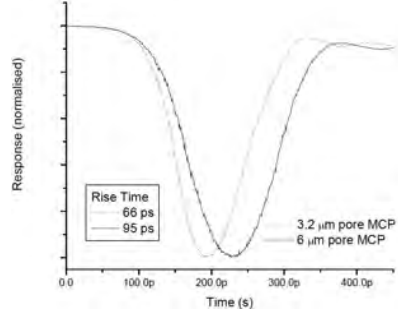

**Figure 3** Photek tube single photon pulses from MCP-PMT [6].

There was an effort in the past 10 years to reach a limit of timing resolution in beam tests using single pixel MCP-PMTs with 6 µm pores. Figure 4 and Table 1 summarizes results. All three tests did not use amplifiers. The CERN RD-51 result by "Picosecond" group is presently the best result of any timing detector in a particle beam [9]. They reached a resolution of ~3.8 ps using a 20 GSa/sec scope employing a CFD algorithm. Tests used two identical MCP-PMTs measuring $\sigma_{double\ MCP-PMT}$.

**Table 1:** Best results achieved using single pixel MCP-PMT detectors ($\sigma_{single\ MCP-PMT} = \sigma_{double\ MCP-PMT}/\sqrt{2}$)

| MCP-PMT Gain | Npe [electrons] | **Total charge [electrons]** | Electronics resolution | Final resolution $\sigma_{single\ MCP-PMT}$ | MCP-PMT type | Reference |
|---|---|---|---|---|---|---|
| ~$2\times10^6$ | ~70 | **~$1.4\times10^8$** | 4.1 ps | 6.2 ps | HPK R3809U-59-11 | [7] |
| ~$10^6$ | ~80 | **~$8\times10^7$** | 2.0 ps | 6.8 ps | Photek 240 | [8] |
| ~$8\times10^4$ | ~44 | **~$3\text{-}4\times10^6$** | 2.2 ps | ~3.8 ps | HPK R3809U-50 | [9] |

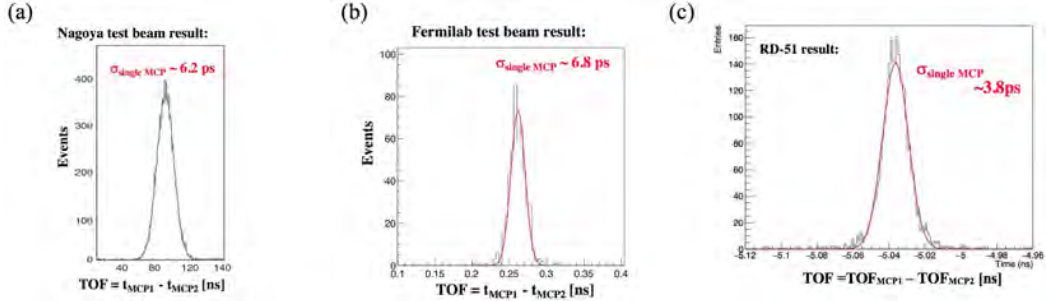

**Figure 4** Comparison of the best timing resolution results with single pixel MCP-PMTs with MIPs in test beams [7,8,9]. No amplifier was used in these tests. See Table 1 for details. (a) and (c) used Hamamatsu R3809U MCP-PMTs, and (b) used Photek-240 MCP-PMT.

The most significant number in Table 1 is the total anode charge per MIP. The problem with running a large total anode charge is a possibility of large after-pulsing rate (caused by ion-feedback), especially when a tube gets older. Figures 5a&b show author's measurement [10] using two old Burle Planacon MCP-PMTs 85013-501. One can see that as long as the total charge stays below a 2-3x10$^6$, the after-pulsing rate is reasonable even with an old Planacon. However, the ion feedback grows exponentially above the charge of 5x10$^6$. One can ask a question if the new MCP-PMTs, ~10 years later, behave better. Figure 5c shows Lehmann's measurements indicating that the new 64-pixel Hamamatsu R13266 (YH0250), with 10μm pores and ALD coating, starts having problems after a total charge of ~2-3x10$^6$ (measurements were done with the single electrons) [11]. On the other hand, a new Planacon Photonis XP85112 (90021008), with 10 μm pores and ALD coating, is OK up a total charge of ~3x10$^6$ [11]. From this point of view, I would question the choice of operating points on figure 4a&b (also ref. [7] and [8] in Table 1), as their total anode charge was extremely large. Such tubes would not last very. On the other hand, the result in figure 4c from Ref.[9] seems reasonable to me, as the total charge was only 3-4x10$^6$.

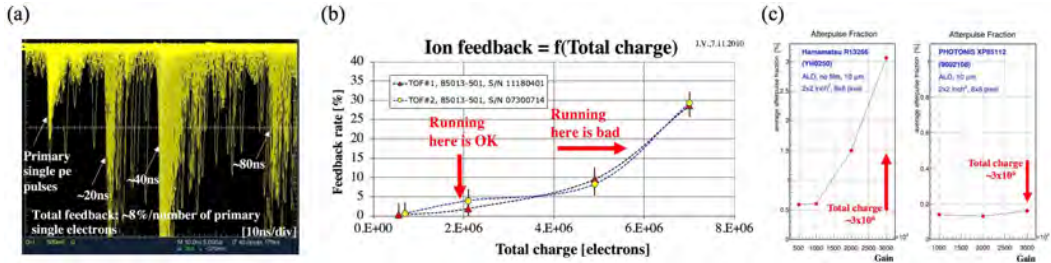

**Figure 5** (a) The after-pulsing rate (or ion feedback) observed in the early versions of Burle Planacon 85013-501 MCP-PMT. The primary signals are single electron primary pulses in this case. Peaks on a storage scope correspond to various ion drift times (H$^+$, H$_2^+$, He$^+$, etc.) [10]. (b) The after-pulsing rate gets worse at higher gain. (c) Lehmann's ion feedback measurement in the latest MCP-PMTs from Hamamatsu (R13266) and Photonis (XP85112) MCP-PMTs [11].

Because of the worry about the excessive after-pulsing rate, I have always argued to run TOF MCP-PMT detectors, operating with large Npe, at low gain with a good low noise fast amplifier. As it is shown on figure 6a, one can obtain a timing resolution of ~14ps for Npe ~40 and ~10ps for Npe ~80 with Hamamatsu C5504-44 amplifier. For Npe ~40, the total charge was only ~8x10$^5$. These results were obtained with several types of electronics (Ortec 9327 CFD, DRS-4 and Wavecatcher) [12,13].

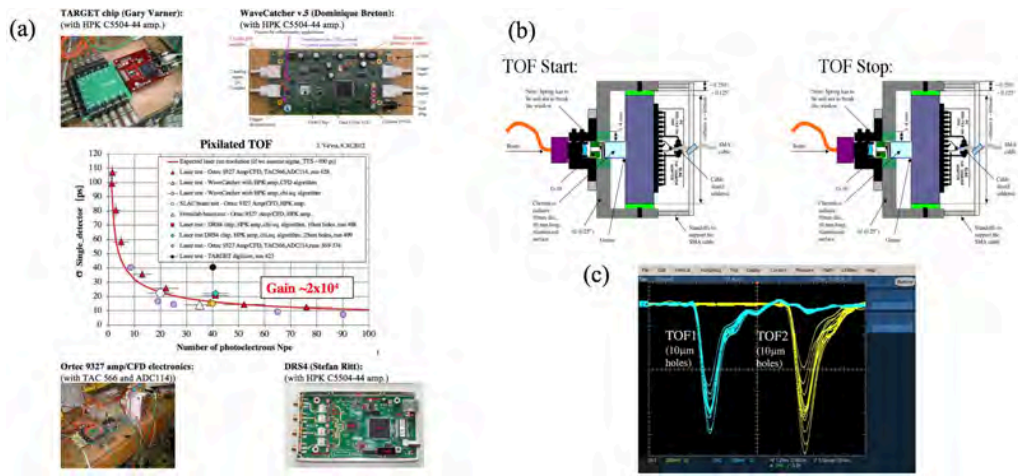

**Figure 6** (a) The timing resolution when operating Planacon 85013-501 MCP-PMT at low gain of ~2x10$^4$ as a function of number of photoelectron [12,13]. One can see that one can achieve a timing resolution of ~15ps for Npe ~40 and ~10ps for Npe ~80. (b) The setup used two Planacon 85013-501 MCP-PMTs (c) Typical MCP-PMT pulses using Hamamatsu C5504-44 1.5 GHz BW amplifier (63x) [12].

Another point we would like to stress that it is not necessary to have a super-fast electronics to obtain a good timing result. One can obtain a good TTS resolution even with a somewhat slower amplifier, if one has a good S/N ratio, and if one tunes the CFD discrimination carefully (every CFD discriminator as an optimum amplitude range, optimum delay and even then, the best results are obtained if one corrects its time determination with an additional pulse height correction). Figure 7 shows timing results with two different amplifiers, the best among five tested. The amplified signal was connected to Phillips 715 CFD and LeCroy 2248 TDC. Planacon 85013-501 MCP-PMT had 10μm pores, operated at a gain of ~10$^6$, and single photoelectron signals were produced by a laser. All pixels of the 64-pixel tube were grounded except the one tested. The quoted resolution was obtained as follows: $\sigma_{TTS} = \sqrt{(32^2 - \sigma_{Laser}^2 - \sigma_{Electronics}^2)}$ ~27 ps. Excellent results proved that the S/N ratio and tuning of CFD operating point are essential to get good timing results [14].

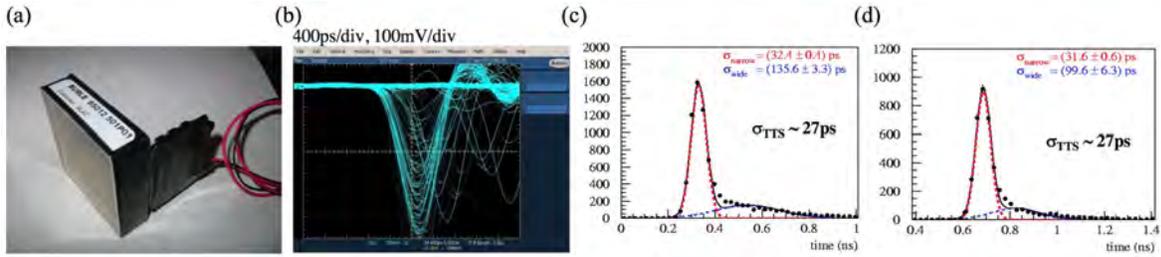

**Figure 7** (a) Planacon 85013-501 with 10μm pores. (b) Single electron pulses with Hamamatsu 63x amplifier C5504-44. Single photon timing resolution with (c) Hamamatsu 1.5 GHz BW amplifier C5594-44 amplifier with 63x gain and (d) Ortec VT120A ~0.4 GHz BW amplifier at 100x gain (200x gain + 6dB). In both cases, Phillips 515 CFD was used [14].

### 3.2 Challenges of multi-pixel MCP-PMTs

Planacon 85011-501, made originally by Burle, was the first multi-pixel MCP-PMT available for physics community and it delivered many good results.[3] For example our group used these tubes for Focusing DIRC prototype, which was the first RICH detector to correct the chromatic error contribution to Cherenkov angle by timing (red photons, which propagate in quartz faster than blue photons, could be tagged by timing in long DIRC bars) [42]. To do this, it was required to measure a single photon timing resolution to 100-200ps. Overall these tubes performed well, on detail level there were some issue.

Although multi-pixel MCP-PMTs are simple devices in principle, there are many challenges how to connect to them. One has to deal with cross-talk between pixels, charge sharing, ringing effects caused by too many hits at the same time, etc. MCP-PMT is inherently a single-ended device referenced to one common ground, which invites a possible pick-up problems. One needs a good RF-shielded box around the device to avoid noise. Early models had also unwanted capacitances, inductances, ground return issues, and low BW connectors, which contributed to cross-talk, pulse shape distortions, ringing, fake hits, etc. However, there is a good news: the latest MCP-PMT tubes are doing better.

Old Burle Planacon:   with SLAC amplifier:   SuperB TOP counter:   Saclay test:   Latest Photonis Planacon:   Latest Photek:

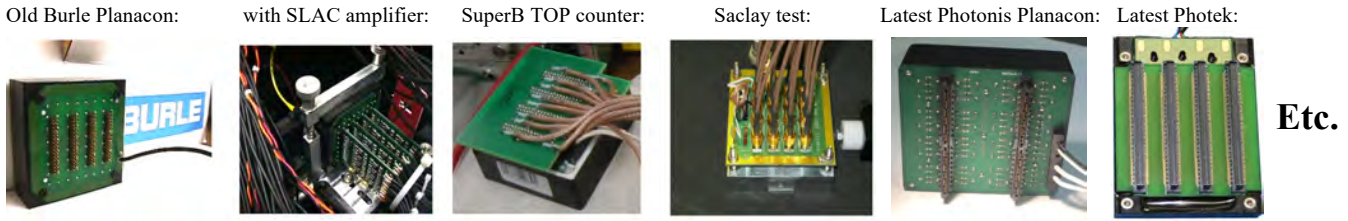

Etc.

**Figure 8** A few examples of connections to Planacon 85011-501 MCP-PMT. Although MCP-PMT is a simple device, connections to it are not simple.

Figure 8 shows a few examples how people connected to MCP-PMT tubes. They all had some pulse quality and cross-talk issues. For example, figure 9 shows author's measurement of the cross talk of an old Burle Planacon 85011-501 MCP-PMT [1,15]. The cross-talk height was typically 3-4% of the primary pulse height and was spread throughout all pixels in a complicated manner. There was a hint that outer pixels were affected most. A possible explanation of the cross-talk, prefering the outer pixels, may have been provided ~12 years later [16]: figure 10 shows a layout of this tube with its resistor divider placed on a PC board around outside boundary. This board may have influenced the cross-talk.

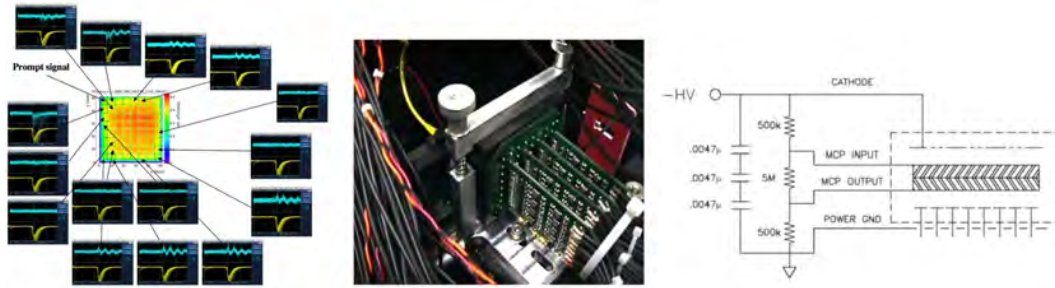

**Figure 9** Author's measurement of the cross-talk of the old Burle Planacon 85011-501 with an internal voltage divider [15]. All pixels were connected to amplifiers with 130x voltage gain. The laser pulse is injected into pixel #1 (upper-left corner). The cross-talk was typically 3-4% of the primary signal and was observed even in far-away pixels in a rather complicated manner. This indicates that the noise is present on inductive ground and distributed around. The picture also shows Elantek amplifier boards used in this test and Burle Co. sketch of grounding.

---

[3] Although Planacon was nominally a 64-pixel device, it was also available as a 1024-pixel device already in 2005; this allowed a user to arrange for his preffered final pixel size by interconnecting small pixels together.

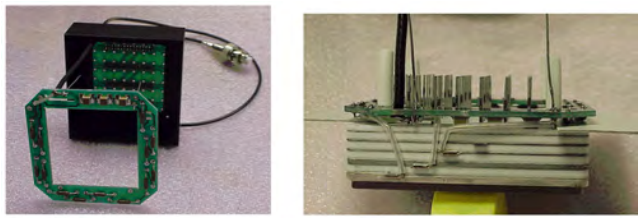

**Figure 10** Outer PC-board layout of old Burle Planacon 85011-501 MCP-PMT with outside resistor divider board [16]. This board influenced the cross-talk.

A consequence of the cross-talk observed in early Planacons, tubes developed "ringing" in a coherent way when exposed to a high multiplicity event when more than 10 pixels/MCP-PMT would fire [17]. This is shown on figure 11a, where one sees that the amplitude of ringing increases with number of photons/event. By the time all pixels fired, the ringing was significant. Such behaviour could create fake hits if the threshold was set too low (for comparison, figure 11b shows that H-8500 MaPMT behaved better under the same conditions.). However, if hit multiplicity was low and threshold set correctly, one could still achieve a very single photon timing resolution with old Burle Planacon, as shown on figure 12 [43]. This was still one of the best timing performance of any large RICH detector system with MCP-PMTs.

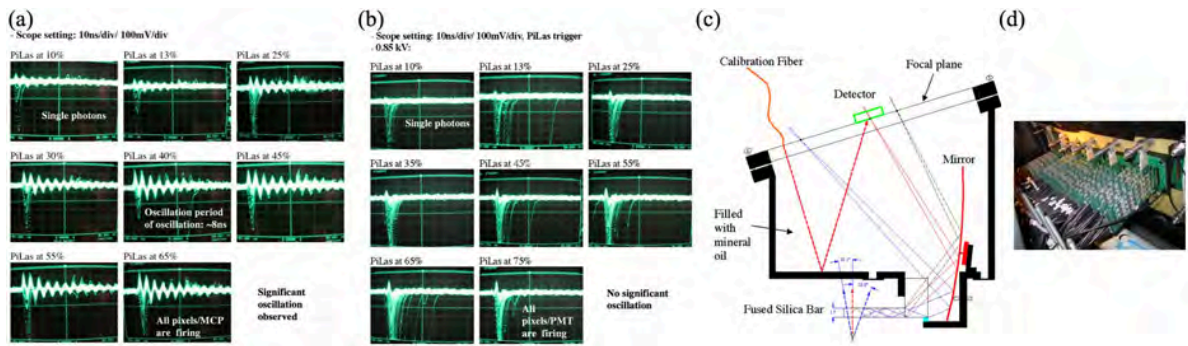

**Figure 11** (a) As a result of cross-talk shown in figure 5, one could create coherent ringing in the entire Burle Planacon 85011-501 MCP-PMT [17]. (b) No significant ringing was observed in Hamamatsu H-8500 MaPMT under the same conditions. (c) This effect was observed in the FDIRC prototype where a laser hit an etched aluminum surface, which acted as a diffuser illuminating all pixels at the same time. (d) Electronics for 320 pixels used in this test [42].

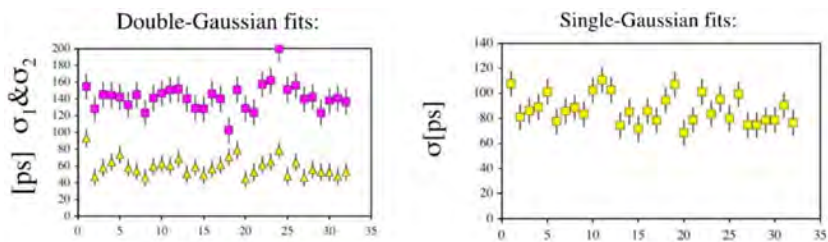

**Figure 12** Singe photon timing resolutions obtained with a PiLas laser and old 64-pixel Burle Planacon MCP-PMT, SLAC amplifier with a gain of 130x and SLAC 32-channel CFD discriminator providing the analog output to a Phillips 7186 TDC (25 ps/count). This was still one of the best timing performance of any large RICH detector system with MCP-PMTs. Burle MCP-PMT used in this test had 25 µm pores with a MCP-to-cathode distance of 0.7 mm (stepped-face MCP-PMT) [43].

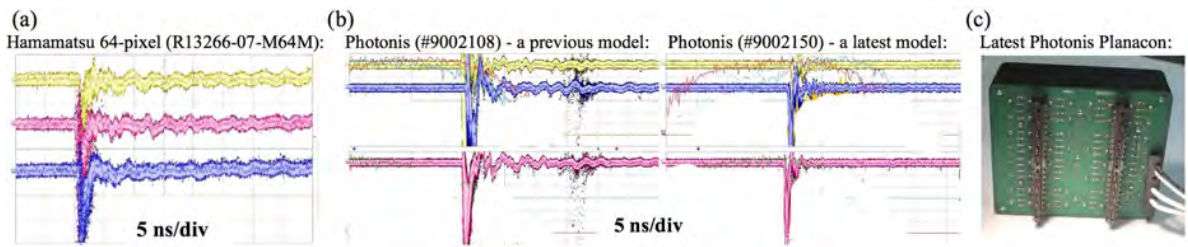

**Figure 13** Lehmann's measurement of ringing of (a) Hamamatsu MCP-PMT when hit by 32 photons at the same time [11] (b) of Photonis tube is subject of an event when all pixels in MCP-PMT have a hit. The most recent Photonis Planacon MCP-PMT model #9002150 is clearly better than the previous models [18] There is clearly an improvement. (c) The latest Planacon MCP-PMT #9002150 with ground plane, improved HV ground return, new connector, smaller anode-ground capacitance [16].

The question is how the new Photonis and Hamamatsu MCP-PMTs behave when exposed to large number of photons/event. This problem is relevant to new experiments, such as Panda DIRCs, TORCH, Belle-II TOP DIRC, which will operate at large rates. Figure 13a shows a significant ringing response of Hamamatsu 64-pixel MCP R13266-07-M64M to ~32 photons/event on average, randomly populating the MCP-PMT face [11]. Figure 13b shows the response of older and very latest Photonis 64-pixel MCP-PMTs [18]. Clearly the latest MCP-PMT #9002150 behaves much better than the old Planacon. The latest MCP-PMT, shown on figure 13c, has a ground plane, an improved HV ground return, a new connector, and a smaller anode-ground capacitance [16].

One should also mention the effort of FIT group at ALICE [19], which had significant impact on understanding the origin of the cross-talk and ringing. Their aim was to reduce 64-pixel of Planacon XP85012 to only four channels, and use it as a TOF detector with four quartz radiators. As figure 14 shows, the cross-talk improved significantly after the Planacon was modified by improving HV ground return and adding a lot of capacitance along MCP-PMT edges.

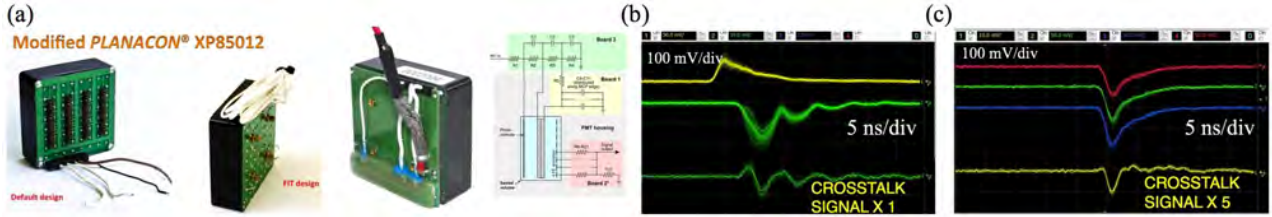

**Figure 14** (a) A modification of 64-pixel Planacon XP85012 to reduce the cross-talk included a use of four SMA connectors, improvement of the HV ground return and increase of a distributed capacitance along MCP-PMT edges. The cross-talk and pulse ringing (b) before and (c) after the modification [19].

Figure 15 shows Burle 1024-pixel Planacon already available in 2005 [44]. SLAC group modified this MCP-PMT into 64-pixel electronics. D. Brasse has used a similar Planacon tube for micro-PET and used all 1024 pixels, each read out by a separate ASIC channel. They achieved a very good point resilution radius of 0.4 mm. In retrospect, if SLAC group would do the same, FDIRC development would take completely different avenue.

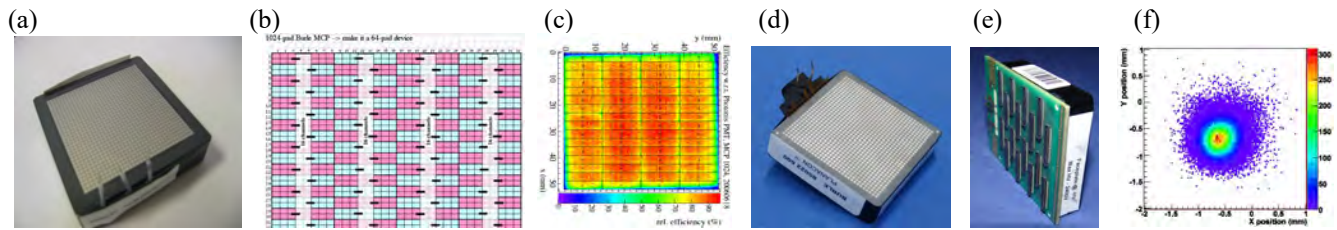

**Figure 15** (a) A 1024-pixel Burle Planacon 85021-600 MCP-PMT with bottom MCP-to-anode distance of 5.2 mm. (b) SLAC has decided to combine small pixels into macro-pixels for 64-channel electronics. (c) The x-y scan using single photons [44]. (d) A similar Planacon 85022-600 MCP-PMT, developed for D.Brasse's micro-PET with a matrix of LYSO crystals, but with bottom MCP-to-anode distance of 3.6 mm. (e) Every pixel was read out into ASIC electronics. (f) The detector achieved the point resolution radius of 0.4 mm [45].

### 3.3 LAPPD Development

Generation-I Incom company MCP-PMTs are now commercially available. These MCP-PMTs have a large size of 20 cm x 20 cm, 25μm pore size, and strip readout. The Incom company has produced about 45 detectors with a typical single photon timing resolution of 60-70ps [20]. The quoted risetime is ~850 ps, therefore one would expect the single photon timing resolution of $\sigma_{time}$ ~60ps for S/N~15 using eq.(1). Figures 16a&b show the detector with its SMA connectors and typical single photon pulses. For many low-rate experiments, this type of detector is an excellent choice. Generation-II MCP-PMTs, with pixel-based readout, will have ceramic body, and capacitive coupling to external PCB board. This concept is still in the R&D stage. Figures 16c&d show a pixel-based detector concept [21] and single photon pulses from pixilated detector obtained either from a direct coupling (red) or from a capacitive coupling.

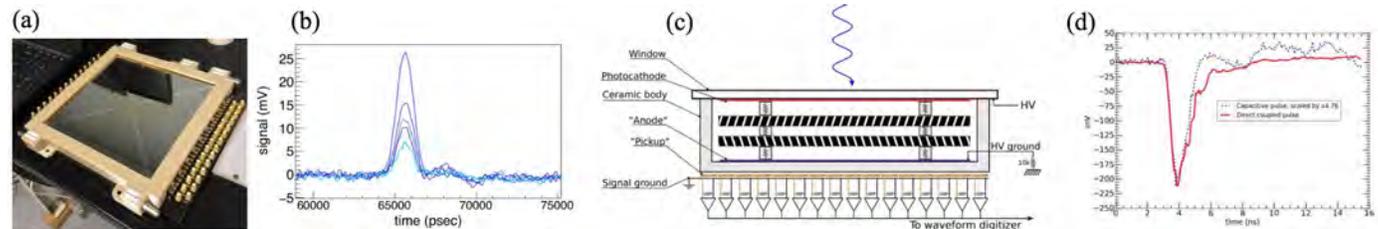

**Figure 16** (a) LAPPD detector with strips and SMA connectors (b) Single photon pulses from the strip detector. (c) Concept of pixilated detector [21]. (d) Single photon pulses from pixilated detector obtained either from direct coupling (red) or from capacitive coupling, normalized to the same amplitude..

### 3.4 MCP-PMT Applications

We will mention three examples from Belle-II, Panda and TORCH.

### Belle-II DIRC (TOP counter)

Belle-II is now taking data, and so they will develop real running experience with 16-pixel R10754-016-M16(N) MCP-PMTs. Figure 17a shows their electronics readout. The readout is based on the IRSX 2.7 GSa/sec waveform digitizer on each pixel. Each pixel is coupled to an amplifier with gain of ~120x with shaping time slowed down to provide 2 samples on the leading edge (figure 17b). The pulse risetime is ~1 ns, and therefore one would expect TTS resolution of $\sigma_{time}$ ~100ps for S/N~10. The bench tests measured TTS resolution of ~83ps (figure 17c). Because of the background at Belle-II, the MCP-PMT gain was lowered to ~$3 \times 10^5$. As a result of this and other effects, the TTS resolution in TOP counter in Belle-II is presently 80-120ps, and the background rate is kept strictly below ~4 MHz/MCP-PMT. Some early non-ALD coated MCP-PMTs may be replaced in 2020 [23].

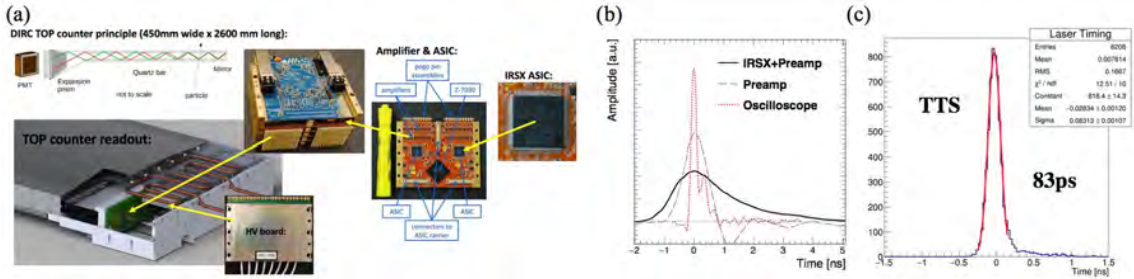

**Figure 17** (a) Waveform digitizing electronics of TOP counter. (b) The MCP-PMT BW had to be slowed down to allow at least 2 samples on leading edge. (c) Single photon (TTS) resolution of ~83ps measured in the lab with the final electronics [22].

### TORCH DIRC at LHCb

TORCH is a TOF detector concept using Cherenkov light to provide charged particle identification up to ~10 GeV/c, which requires a timing resolution of ~15ps/track. The TORCH fused silica radiator is covering a large area, with focusing optics and MCP-PMT readout located at two edges. Photons propagate in these plates up to path lengths of 5-8 meters. The detector design concept is taking advantage of the fact that there is a very long time of flight path between the vertex and TORCH detector in LHCb [24,25]. In theory, if all MCP-PMT time offsets are aligned, $t_o$ start time negligable, geometry of the detector is understood, the chromatic broadening corrected out by pixels, and number of photons Npe ~25 detected, the TORCH detector could achieve $\sigma_{TOF} \sim \sigma_{TTS}/\sqrt{(Npe)} \sim 70/\sqrt{25} \sim 14$ ps in principle. In my view, this will not be easy in practice.

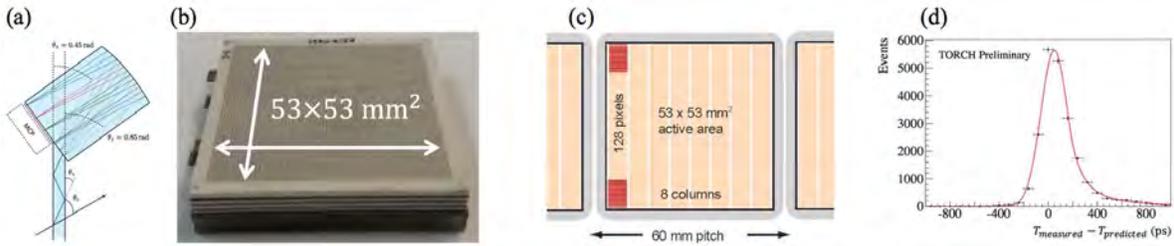

**Figure 18** (a) Optics used at the end of fused silica plates. (b) MCP-PMT used in TORCH tests made by Photek, Ltd [26]. (c) MCP-PMT has 8 columns each with 128 strips with pitch of 0.414 mm. (d) The best single photon timing resolution achieved was 88.8 ± 1.3 ps (narrow portion of shape) for the shortest photon path length in their TORCH prototype, so there is an additional contribution to the TTS resolution [27]. The prototype used NINO ASIC [5].

Each Photek MCP-PMT has 8 columns, each with 128 strips with pitch of 0.414 mm [26]. Initial prototype tests indicate a single photon timing resolution of 80-100 ps/photon at a gain of ~$10^6$ (see figure 18c). The prototype used the NINO charge amplifier/discriminator ASIC with time-over-threshold (TOT) pulse height correction, coupled to HPTDC. Conditions at LHCb will be severe, with predicted rates up to 10-40 MHz/cm$^2$, and total anode charge doses up to ~5C/cm$^2$. So far, aging tests with Phase-I Photek MCP-PMT indicate a "good MCP-PMT QE behavior" up to ~3C/cm$^2$. I was not able to obtain raw pulses from the prototype.

### Endcap Panda DIRC

Panda Endcap DIRC [28] has many similar features to TORCH. However, since it does not have an advantage of long particle path lenghts like TORCH, it will rely on pixel-based analysis of the Cerenkov angle with some help of timing. Figure 19 shows a focusing optics located along edges of large fused silica plates with MCP-PMT and electronics. The experiment plans to use Photonis XP85132-S-MD3 MCP-PMT with 3 rows x 100 strip configuration, 0.4 mm x 17 mm anode pads, and MCP-anode gap equal to 0.625 mm [28]. Figure 19c shows a very good position resolution when MCP-PMT was placed in a small magnetic field of 0.1T [29]. The electronics is based on TOFPET ASIC, which was developed initially for the Time-of-Flight Positron-Electron

Tomography, where a TTS resolution of ~100ps was achieved with SiPMTs (Hamamatsu S13361) – see figure 19d. The main advantage of TOFPET ASIC is that it has low cost, low mass, small power consumption and it is radiation hard. However, the initial design of TOFPET was made for positive pulses, such as in SiPMTs; the design has to be modified for negative pulses, which is being done presently.

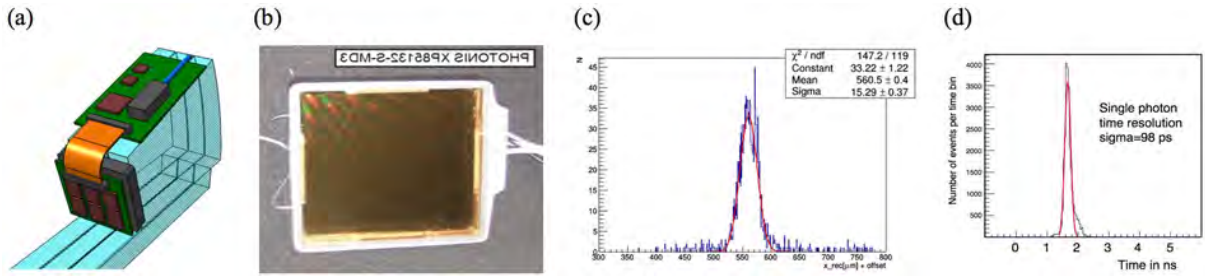

**Figure 19** (a) Optics used at the end of fused silica plates with MCP-PMT and electronics layout. (b) The experiment plans to use Photonis XP85132-S-MD3 MCP-PMT with 3 rows x 100 strip configuration, 0.4 mm x 17 mm anode pads, and MCP-anode gap = 0.625 mm. The tube's anode strips are grounded by electronics, i.e., tube does not have a ground plane [16]. (c) A very good position resolution was achieved when MCP-PMT was placed in small magnetic field of 0.1T [29]. (d) TOFPET electronics has achieved previously a single photon resolution of ~100 ps with SiPMTs.

## 4. Silicon detectors

Many examples were discussed in Ref.[1]. Here we provide only two important latest developments.

SiPMTs are known to suffer from the neutron damage. As figure 20a shows, their damage is significant above a neutron dose of ~$10^{11}$ $n_{eq}/cm^2$, if detector is operated at room temperature [46]. Excessive noise prevents their use for typical RICH detectors operating at room temperature. However, if SiPMTs are cooled the noise level is significantly improved, as figure 20b indicates [47]. It was found that high energy protons and neutrons produce the most damage. Damage from thermal neutrons is observed only at high doses. Gammas produce comparatively lower damage. Lower temperature can reduce noise rate caused by the neutron damage. All SiPMTs, even those irradiated up to $10^{14}$ $n_{eq}/cm^2$, are "usable" at liquid nitrogen temperature. Operation for RICH detectors in single photon regime at $10^{11} n_{eq}/cm^2$ and –30 °C is possible [47]. Figure 21 shows application of SiPMTs for focusing ARICH for EIC. Cherenkov ring, obtained in test beam, clearly shows a noise improvement at -30°C [48].

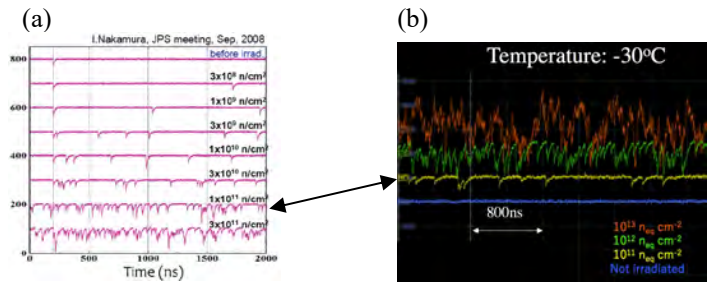

**Figure 20** (a) SiPMT noise level as a function of neutron dose [46]. (b) SiPMT noise level at -30°C as a function of neutron dose [47].

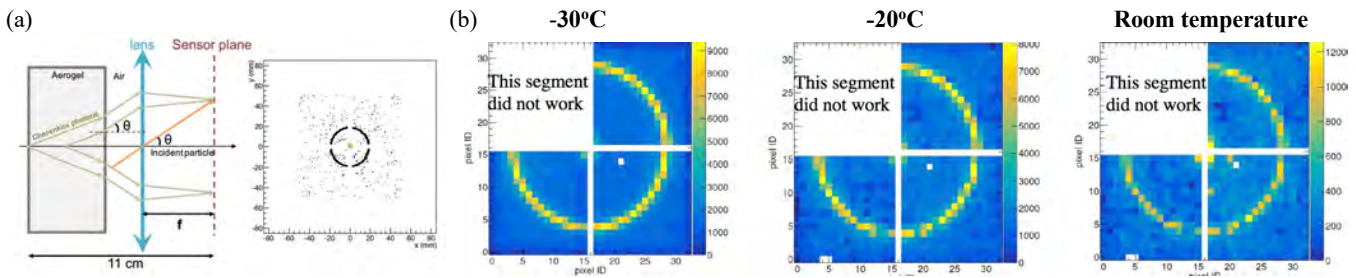

**Figure 21** (a) Principle of ARICH with Fresnel lens focusing [48]. (b) Cherenkov ring images taken at -30°C, -20°C and room temperature [48].

There is a strong motivation to develop new Si 3D-detecors. The idea is to connect charged tracks to the correct production vertices, using position resolution and timing resolution. The emerging technology of choice is low gain avalanche diode (LGAD) Si detectors. Figure 22a shows a principle of LGAD [33]. The proof of principle was provided in a test beam, where proponents measured $\sigma_{time}$ ~34 ps for a single sensor, and ~16 ps with a tandem of three identical sensors (figure 22b,c) [34]. Each sensor size was 1.3mm x 1.3mm x 45 μm-thick, equipped with ~1.5 MHz BW amplifier and a 20 GSa/sec oscilloscope readout, providing 50 ps sampling. Figure 22b indicates a risetime of ~400ps and a S/N ratio of ~20; therefore, using formula (1), one would expect $\sigma_{time}$ ~ $t_{risetime}$ /(S/N) ~20 ps.

ATLAS and CMS collaborations are developing new ASICs, which will be located on each sensor. An example is shown on figure 22d [35]. Figure 22e shows endcap application for ATLAS. It is now considered for the High-Granularity Timing Detector (HGTD) [35]. Presently proponents tune various designs, figure 22f shows one example, which has a 90% fill factor. In dead region, position and time are determined by amplitude-weighted centroid using four pads, as indicated on figure 22g. One should point out that the expected total doses at ATLAS are extremely challenging (~4×10$^{15}$ n$_{eq}$/cm$^2$ and ~4.1 MGy = ~410 Mrads), however, tests up to this moment indicate that proponents are close to this goal. For comparison, BaBar DIRC optical components were tested up to ~10-20 krads only. Bench tests achieved very good timing and position resolution results using a laser (σ ~ 10's of ps & 10's of μm).

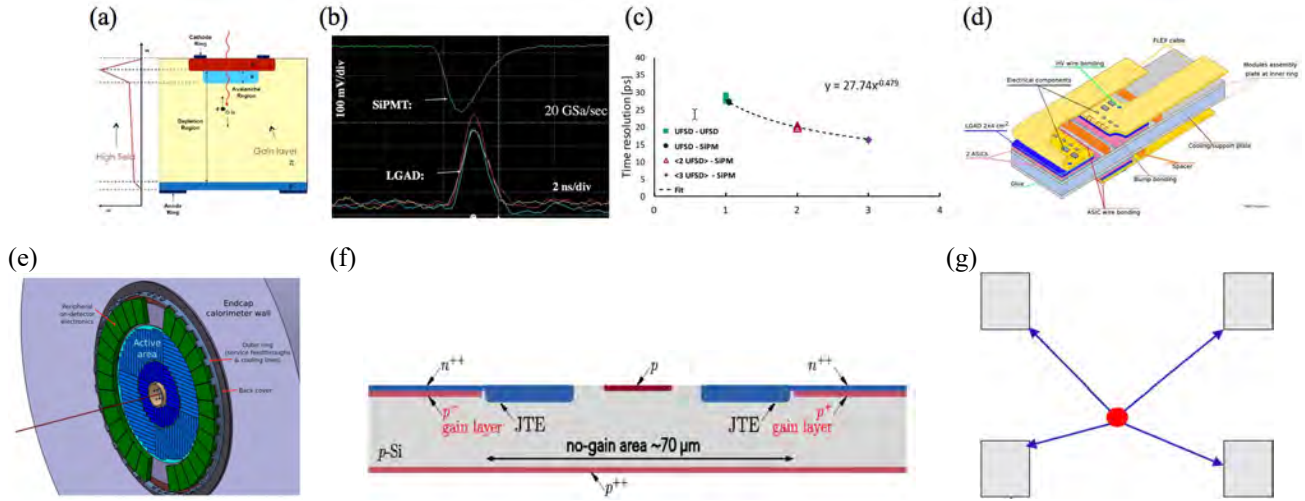

**Figure 22** (a) Principle of the LGAD detector [33]. (b) Pulses from LGAD detector, compared to SiPMT in a test beam [34]. (c) Resolution as a function of number of layers in tandem of identical detectors, obtained in a test beam [34]. (d) A proposed LGAD tile structure for ATLAS HGTD detector [35]. (e) ATLAS UFSD Endcap (12 cm < r < 60cm) with 7888 sensor modules. (f) Present design have a region of no gain (pitch: 1.3 mm, gap: ~ 70mm fill factor: ~90%). (g) In the gap with no gain, position and time are determined by amplitude-weighted centroid using four pads.

## 5. Gaseous detectors

The last example is the gaseous Micromegas and one is the ATLAS LGAD detector, shown on figure 23. The main attraction of Micromegas detectors [30] is that they can be made in our labs and therefore one controls its basic design parameters. They use typically CsI photocathodes, deposited on a 3mm-thick MgF$_2$ window. Recent results are very impressive. Micromegas has achieved a timing resolution of ~24 ps with 150 GeV/c muons, and ~76 ps resolution with single photoelectrons in laser tests [31]. These results were achieved with a mean number of photoelectrons of ~10 per muon. Recently, new Diamond-Like Carbon (DLC) photocathodes are under investigation [32]. These photocathode are expected to be more radiation hard compared to CsI. So far, their QE is about ~3x lower than what one gets from CsI [40].

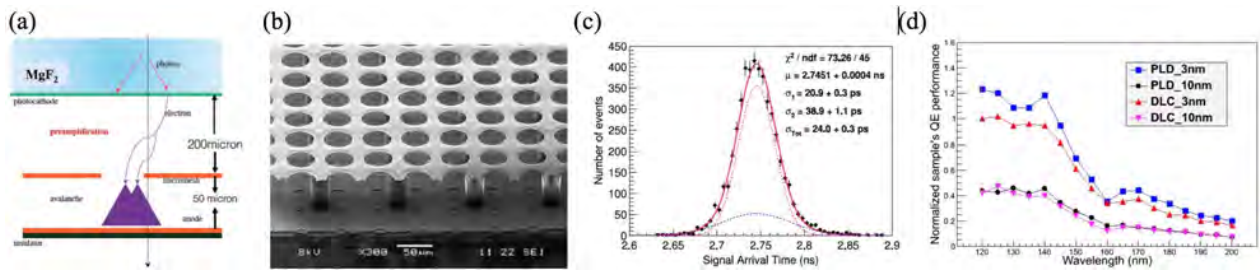

**Figure 23** (a) Concept of the Micromegas gaseous detector. (b) A detail of the gain region of the Micromegas. (c) Timing resolution obtained in muon beam with 3mm MgF$_2$ widow and CsI photocathode [31]. (d) A measured relative QE of new Diamond-Like Carbon (DLC) photocathodes [32]; one needs another factor of 3x in absolute QE to get a similar result as from the CsI photocathode [40].

## 6. Summary of timing resolution, rate capability and aging tests

It is useful to summarize the present status of MCP-PMT capability. Figure 24a shows Panda measurements of MCP-PMT rate capabilities [37]. Typically, the older MCP-PMTs show stable operation to ~200-300 kHz/cm$^2$ of single photons at gain of 10$^6$. Some of the latest MCP-PMTs can push the single photon rate up to ~10 MHz/cm$^2$ at this gain. Figures 1.4.2b,c show a stability of QE as a function of total anode charge [11,38]. Some of the latest Photonis and Hamamatsu ALD-coated MCP-PMTs seem to be stable up to ~20 C/cm$^2$. This represents a remarkable progress in the past ~10 years. However, it has to be proven in a real experiment.

Table 2 shows a summary of rate and aging measurements from other detectors. It also lists obtained timing resolutions.

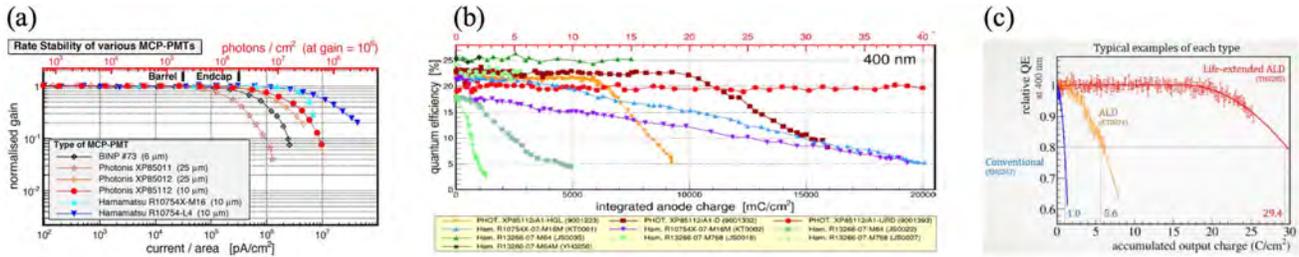

**Figure 24** (a) Measured rate capability of MCP-PMTs [37]. (b) Measured photocathode deterioration as a function of accumulated anode charge. Some latest Photonics MCP-PMTs can reach anode charges of ~20 C/cm$^2$ [11]. (c) Measurement of QE aging of some Hamamatsu MCP-PMTs indicate that the latest tubes can reach ~20C/cm$^2$ [38].

**Table 2:** Timing resolution, maximum rate capability and anode charge doses, obtained in beam tests or laser tests, (some numbers are either not known to author presently or not measured).

| Detector | Experiment or beam test | Maximum rate | Maximum anode charge dose | Timing resolution | Ref. |
|---|---|---|---|---|---|
| MRPC presently | ALICE | ~500 Hz/cm$^2$ *** (tracks) | - | ~60 ps/track (present)*** | [4] |
| MRPC after upgrade | ALICE | Plan: ~50 kHz/cm$^2$ ** (tracks) | - | Plan: ~20 ps/track | [4] |
| MCP-PMT | Beam test | - | - | < 10 ps/track * | [7,8,9] |
| MCP-PMT | Laser test | - | - | ~27 ps/photon * | [14] |
| MCP-PMT | PANDA Barrel test | 10 MHz/cm$^2$ * (laser) | ~20 C/cm$^2$ * | - | [11] |
| MCP-PMT | Panda Endcap | ~1 MHz/cm$^2$ ** (photons) | - | - | [28] |
| MCP-PMT | TORCH test | - | 3-4 C/cm$^2$ * | ~90 ps/photon * | [27] |
| MCP-PMT | TORCH | 10-40 MHz/cm$^2$ ** (photons) | 5 C/cm$^2$ ** | ~70 ps/photon ** | [24-27] |
| MCP-PMT | Belle-II | < 4MHz/MCP *** (photons) | - | 80-120 ps/photon*** | [23] |
| Low gain AD | ATLAS test | ~40 MHz/cm$^2$ ** (tracks) | - | ~ 34 ps/track/single sensor * | [34,35] |
| Medium gain AD | Beam test | - | | < 18 ps/track * | [39] |
| Si PIN diode (no gain) | Beam test (electrons) | - | | ~23 ps/32 GeV e$^-$ | [8] |
| SiPMT (high gain) | Beam test – quartz rad. | - | < 10$^{10}$ neutrons/cm$^2$ | ~ 13 ps/track * | [8] |
| SiPMT (high gain) | Beam test - scint. tiles | - | < 10$^{10}$ neutrons/cm$^2$ | < 75 ps/track * | [41] |
| Diamond (no gain) | TOTEM | ~3 MHz/cm$^2$ * (tracks) | - | ~ 90 ps/track/single sensor * | [36] |
| Micromegas | Beam test | ~100 Hz/cm$^2$ * (tracks) | - | ~24 ps/track * | [31,32,40] |
| Micromegas | Laser test | ~50 kHz/cm$^2$ * (laser test) | - | ~76 ps/photon * | [31,32,40] |

\* Measured in a test
\*\* Expect in the final experiment
\*\*\* Status of the present experiment

## 7. Summary

- MCP-PMT detectors timing resolution limit has been pushed to ~3.8 ps in beam tests using single pixel devices.
- MCP-PMTs are fastest detectors available presently. This speed creates challenges for multi-pixel applications. Although some of these problems were pointed out already ~15 years ago, it is finally now that we see a progress in design of these detectors. There is a significant progress in MCP-PMT aging and rate properties.
- There is a large effort, driven mainly by LHC, to develop a Si-tracker providing a time resolution at a 20-30 ps level per track, based on low gain avalanche detectors (LGAD). Expected total radiation doses at ATLAS are extremely challenging (~4×10$^{15}$ n$_{eq}$/cm$^2$ and ~4.1 MGy = ~410 Mrads.
- Micromegas gaseous detectors reached a timing resolution of ~24 ps. This is amazing achievement, not expected a few years ago. The main advantage of this type of detector is that one can create it in our labs. New Diamond-Like Carbon (DLC) photocathodes may be a significant step to improve their aging behavior. They still need a factor of 3 improvement in QE to compete with the CsI photocathodes.
- However, most results in this paper were obtained in small tests, usually with single-pixels. The timing resolution will be worse in large complex multi-pixel detector systems with a real background.

## ACKNOWLEDGEMENTS

I would like to thank C. Williams, A. Lehmann, S. White, H. Sadrozinski, K. Nishimura, Y. Giomataris, Jeff DeFazio and E. Schyns for contributions to this talk.